\documentclass[superscriptaddress,twocolumn,showpacs,prl,floatfix]{revtex4}

\usepackage{color}
\usepackage{tabularx}
\usepackage{epsfig}
\usepackage{amsmath}
\usepackage{graphicx}

\begin{document}

\title{Critical scaling to infinite temperature}

\author{P. H.~Lundow}
\affiliation {Department of Theoretical Physics, Kungliga Tekniska h\"ogskolan, SE-106 91 Stockholm, Sweden}

\author{I. A.~Campbell}
\affiliation{Laboratoire Charles Coulomb,
  Universit\'e Montpellier II, 34095 Montpellier, France}

\begin{abstract}
Three dimensional Ising model ferromagnets on different lattices
with nearest neighbor interactions, and on simple cubic lattices with
equivalent interactions out to further neighbors, are studied
numerically.  The susceptibility data for all these systems are analyzed using the critical Renormalization Group Theory formalism over the entire
temperature range above $T_c$ with an appropriate choice of scaling variable and scaling expressions. Representative experimental data on a metallic ferromagnet (Ni) and an elementary fluid (Xe) are interpreted in the same manner so as to estimate effective coordination numbers.
\end{abstract}


\maketitle

In the very extensive studies which have been devoted to critical
phenomena attention has understandably been mainly concentrated on the
regime in the immediate neighborhood of the critical temperature; data are
analyzed using the leading terms in the Renormalization Group Theory
(RGT) formalism. It is widely considered that as correction terms
proliferate outside this narrow "critical region" they always lead ultimately towards Gaussian fixed point mean-field-like behavior
above a temperature $T_g$ determined by the Ginzburg criterion
\cite{ginzburg:60}. This criterion expresses a crossover from a fluctuation dominated critical regime to a high temperature Landau regime.
Sophisticated theoretical and numerical studies have been made of this
crossover
\cite{riedel:69,luijten:97,pelissetto:98,luijten:99,anisimov:99,garrabos:08} in particular in the context of long range interactions. Experimental
data on fluids, where effective interactions are expected to be long
range, have been interpreted on this basis
\cite{anisimov:95,kim:03,sengers:09}.

Here we discuss from a different perspective numerical data on various
3d Ising systems, in particular models where the interactions extend
beyond nearest neighbor. We conclude that with appropriate observables
and choice of scaling variable, for this family of models at least
there is no need to invoke a crossover or the Ginzburg criterion. The
data can be convincingly interpreted using the rigorous critical RGT formalism
over the whole temperature range from $T_c$ to infinity. Applying Occam's razor, this approach is more economical conceptually as it requires only the extended critical analysis but no separate analysis linked to a crossover.

We discuss representative experimental data on a ferromagnet and on a fluid, showing that they too can be analyzed in a transparent manner using the same approach and without invoking crossovers. We obtain
quantitative estimates of the effective coordination number for the metallic ferromagnet Ni and for the elementary fluid Xe.

Before RGT it was already established \cite{gammel:63,fisher:67,stanley:71} that in the fundamental scaling law for ferromagnetics the
response parameter is the "reduced" susceptibility (see
\cite{stanley:71} for definitions) :
\begin{equation}
\chi(\beta) = \langle \left(m - \langle m\rangle\right)^2\rangle =\sum_{i,j}\langle S_{i}\cdot S_{j}\rangle =
\chi_{T}(\beta)/\chi_{0}(\beta)
\end{equation}
where the thermodynamic susceptibility
\begin{equation}
\chi_{T}(\beta) \equiv [\partial{m}/\partial{H}]_{H \to 0}
\end{equation}
is normalized by the free spin susceptibility $\chi_{0}(\beta) \propto
\beta$. (As usual we will set the interaction strength $J$ to $1$ and
write $\beta \equiv 1/kT$).  The critical behavior is written
\cite{stanley:71,butera:02,gartenhaus:88,orrick:01}
\begin{equation}
\chi(\beta) \equiv T\chi_{T}(\beta) \propto \epsilon^{-\gamma}\left(1
+ \cdots\right)
\end{equation}
where $\epsilon$ is an appropriately normalized scaling variable
depending linearly on $(T - T_c)$ close to $T_c$.

The thermodynamic ideal lattice gas analogue to $\chi_{T}$ is the isothermal compressibility on the critical isochore
$K_{T}=-(\partial{V}/\partial{p})/V$ so by strict analogy to the ferromagnetic case the parameter which should be scaled
\cite{stanley:71} is the compressibility normalized by the ideal gas
compressibility $\langle(N-\langle N \rangle)^2 \rangle/\langle N
\rangle$ where $N$ is the total number of particles, i.e.
\begin{equation}
TK_{T} \propto \epsilon^{-\gamma}[1 + \cdots]
\end{equation}
Because the gas-liquid order parameter is a scalar the
fluid transition belongs to the Ising universality class
\cite{kadanoff:71}. Careful experimental measurements and analyses made over many years (see \cite{sengers:09}) have shown that the asymptotic fluid critical exponents are indeed those of the short range Ising
universality class. The real situation is however more complicated than in the magnetic case because of the departure from vapor-liquid symmetry in real fluids \cite{fisher:00,orkoulas:01,sengers:09}; the fluid "susceptibility" is defined in Ref.~\cite{sengers:09}.

In work based on the high temperature series expansion (HTSE) theory which was already firmly established in the 1950s, the critical scaling variable is taken to be either  $\epsilon =\tau = [1-\beta/\beta_c]$ or $\epsilon = [1-\tanh(\beta)/\tanh(\beta_c)]$
\cite{gammel:63,fisher:67,butera:02}. Appropriate high temperature series are written rigorously as sums of terms where exact factors multiply successive powers of $\epsilon$. However since the introduction of RGT, scaling expressions are often written in terms of $\epsilon = t = [(T-T_c)/T_c]$. This is just the simplest linear convention, but other scaling variables including $\tau$ are just as legitimate as $t$ in the region very close to $T_c$. (For instance in the special case of the square lattice Ising model an extremely sophisticated analysis uses $(1/\sinh(2\beta)-\sinh(2\beta))/2$ as the scaling variable \cite{orrick:01}). In addition $\tau$ has obvious practical advantages in the temperature region well above $T_c$ because its high temperature limit is $1$ (and not infinity as is the case for $t$).

Systematic analyses of high temperature
numerical data of nearest neighbor and long range interaction models in the Ising universality class
\cite{anisimov:95,luijten:97,garrabos:08,chamberlin:09}, have been carried out with an
effective temperature dependent susceptibility exponent defined as
\begin{equation}
\gamma_{t}(t) = -\partial{\log \chi_{T}(\beta)}/\partial{\log t}
\label{gammatdef}
\end{equation}
following the phenomenological expression of Kouvel and Fisher \cite{kouvel:64}. In the high temperature limit for any spin model $\chi_{T} \to \beta$ and $t \to T$ so
$\gamma_{t}(t)$ will automatically tend to $1$ at high temperatures. There will necessarily be a crossover at an intermediate temperature where $\gamma_{t}(t)$ passes from the critical $\gamma$ to $1$. However, far from criticality  the choice of variables is vital. Thus if the reduced susceptibility parameter $\chi(\beta)$ rather than $\chi_{T}(\beta)$ had been used for the definition of $\gamma_{t}(t)$ in Ref.~\cite{luijten:97} and following work, the effective exponent would have tended to $0$ at high temperature, not to $1$.

The full formal RGT Wegner scaling expression for $\chi(\tau)$ (rather than $\chi_{T}(t)$) in the thermodynamic limit including confluent and
analytic correction plus background terms is written rigorously using $\tau$ as
\cite{wegner:72,aharony:83,gartenhaus:88,butera:02}
\begin{multline}
\chi(\tau)= T\chi_{T}(\tau) = C_{\chi}\tau^{-\gamma} [1
  + a_{\chi}\tau^{\theta}F_{a}(\tau)+ b_{\chi}\tau F_{b}(\tau)   \\
  + c_{\chi}\tau^{(1-\alpha)\gamma}F_{c}(\tau) + d_{\chi}\tau^{\gamma}F_{d}(\tau)
  + a_{2,\chi}\tau^{\theta_{2}}F_{2}(\tau) +\cdots]
  \label{chiwegner}
\end{multline}
where $\gamma, \alpha$, the confluent correction
exponents $\theta_{i}$, and certain amplitude ratios are universal but the  amplitudes themselves are not
universal; the $F_{i}$ are infinite analytic series in $\tau$
normalized to $1$ at $\tau=0$. Because $\tau \to 1$ as $T \to \infty$
these developments remain well behaved at all temperatures above $T_c$, whereas because $t$
diverges as $T$ it is obviously very awkward to extend to high temperatures the analogous expression written in terms of $t$.

The temperature dependent effective susceptibility exponent defined in terms of
$\tau$ and $\chi(\tau)$ in
Refs.~\cite{fahnle:84,orkoulas:00,butera:02,campbell:07,campbell:11} is
\begin{equation}
\gamma_{\mathrm{eff}}(\tau) = -\partial{\log\chi(\tau)}/\partial{\log(\tau)}
\label{gammadef}
\end{equation}
with the equivalence
\begin{equation}
\gamma_{t}= \gamma_{\mathrm{eff}}(1-\tau)+\tau
\label{gammat}
\end{equation}
Very close to $T_c$ the two effective $\gamma$ parameters are indistinguishable but they have quite different properties as soon as the temperature difference increases. (Historically it is of interest to note that the equation \cite{gammel:63}
\begin{equation}
1/\chi_{T}= (1/C)T(1-(T_c/T))^{\gamma}f(T_c/T)
\label{chigammel}
\end{equation}
cited explicitly by Kouvel and Fisher \cite{kouvel:64} to justify their analysis in the vicinity of $T_c$ is precisely of the form of Eq.~\ref{chiwegner}, with
$\tau$ as the scaling variable).

At first sight the sets of infinite series of corrections in Eq.~\ref{chiwegner} appear rather forbidding.  However, from inspection of
$S=1/2$ high temperature series expansions (HTSE) (see for instance \cite{butera:02}) there are exact
closure rules at infinite temperature : $C_{\chi}(1 + a_{\chi} +
  \cdots) \equiv 1$ and $\gamma_{\mathrm{eff}}(1) \equiv z\beta_c$
\cite{fahnle:84} where $z$ is the coordination number and the
$(\cdots)$ represent the exact sum of all the higher order terms in Eq.~\ref{chiwegner} evaluated with $\tau$ set equal to $1$. It turns out that for the 3d Ising models which we will discuss explicitly, over a wide temperature region above $T_c$ the leading [confluent] Wegner correction term dominates. As a convenient approximation all the remaining terms can be collected together into a single weak effective correction term $k_{\chi}\tau^{\lambda_{\chi}}$, giving a compact approximate expression which can be used to fit the data over the entire temperature range above $T_c$ :
\begin{equation}
\chi(\tau)\tau^{\gamma} = C_{\chi}(1 + a_{\chi}\tau^{\theta}
  + k_{\chi}\tau^{\lambda_{\chi}})
  \label{chitau}
\end{equation}
so
\begin{equation}
\gamma_{\mathrm{eff}}(\tau) = \gamma -(a_{\chi}\theta\tau^{\theta} +
k_{\chi}\lambda_{\chi}\tau^{\lambda_{\chi}})/(1 + a_{\chi}\tau^{\theta} +
k_{\chi}\tau^{\lambda_{\chi}})
\end{equation}
The closure rules then become
\begin{equation}
C_{\chi}(1+ a_{\chi} + k_{\chi}) = 1
\end{equation}
and
\begin{equation}
\gamma - (a_{\chi}\theta + k_{\chi}\lambda_{\chi})C_{\chi} = z\beta_c
\end{equation}
Once the strictly defined critical amplitudes $C_{\chi}$ and $a_{\chi}$ are estimated for any particular model from data at temperatures
close to criticality, $k_{\chi}$ and $\lambda_{\chi}$ are fixed also from the closure conditions,
so the entire temperature dependencies of $\chi(\tau)$ and of
$\gamma_{\mathrm{eff}}(\tau)$ are determined. (An expansion of Eq. \ref{chitau} to include further explicit Wegner terms is possible when HTSE data and higher order correction exponent values are available).

While not rigorous except in the limits $\tau \to 0$ and $\tau \to 1$, this {\it ansatz} as it stands already gives a representation of the true behavior of $\chi(\tau)$ which turns out to be accurate to the $10^{-3}$ level over the whole temperature range above $T_c$ for all the models we have studied. If $\gamma_{\mathrm{eff}}(\tau)$ is transposed to  $\gamma_{t}$ through Eq.~\ref{gammat} a crossover behavior results (see Fig. 4); similar data have been analyzed using an approximant containing an implicit crossover function \cite{anisimov:99,kim:03}.

The temperature dependent susceptibility $\chi(\beta,L)$ was
evaluated on diamond, sc, bcc and fcc lattices with nearest neighbor
interactions, see \cite{haggkvist:07,lundow:09,campbell:11} where the
numerical techniques are described, and on sc lattices with equivalent
interactions up to second, third, fourth or fifth neighbor following
\cite{domb:66,luijten:97,orkoulas:00}. The coordination numbers are $z
= 4, 6, 8, 12, 18, 26, 32$ and $56$ respectively. As all these models are
in the 3d short range interaction Ising universality class, in the analysis  below the numerical values for the universal 3d Ising exponents were fixed at $\gamma=1.2371$, $\theta=0.50$ and $\nu=0.630$
\cite{pelissetto:02,deng:03}. The critical inverse temperatures
$\beta_c$ for the various models were evaluated from the present data
using the Binder cumulant $g(\beta,L)$ and the parameter $W(\beta,L)$
introduced in \cite{lundow:10}. The $\beta_c$ values obtained from the
finite size scaling analysis are in full agreement with previous
estimates, in particular those of Ref.~\cite{luijten:99} for the equivalent interaction models.  The raw normalized reduced susceptibilities
$\chi(\tau,L)\tau^{\gamma}$ as a function of $\tau^{\theta}$ for
the $z26$ (nearest, second nearest, and third nearest equivalent neighbors) sc Ising model at different sizes $L$ are shown as an example in Fig. 1. The envelope curve which can be seen by inspection corresponds to the thermodynamic limit (effectively infinite size) behavior. Data for each of the other models have qualitatively similar appearance (see \cite{campbell:11} for the nearest neighbor sc model). The susceptibility results for infinite $L$ (extrapolated for $\tau$ close to zero) for the various coordination numbers $z$ are exhibited in Fig. 2 in the form of plots of
$\chi(\tau)\tau^{\gamma}$ against $\tau^{\theta}$. The effective exponents
$\gamma_{\mathrm{eff}}(\tau)$ derived from these data are shown in Fig. 3.
\begin{figure}
  \includegraphics[width=3.5in]{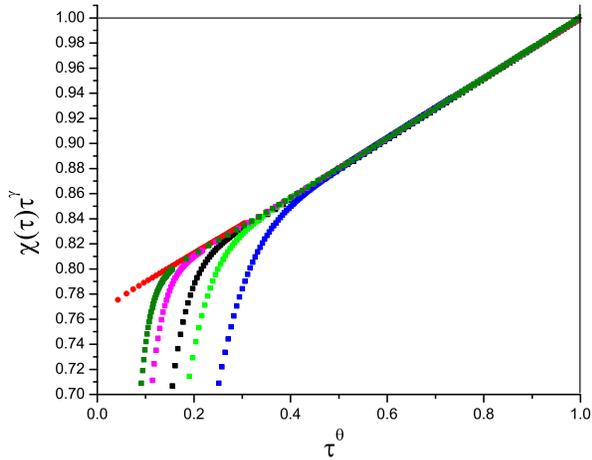}
  \caption{(Color online) The normalized reduced susceptibility
    $\chi(\tau)\tau^{\gamma}$ as a function of $\tau^{\theta}$ for
    the sc $z26$ (nearest, second nearest, and third nearest equivalent neighbors) Ising model. Lattice sizes $L= 32, 24, 16, 12, 8$
top to bottom (olive, pink, black, green, blue). The $L > \xi(L,\beta)$ size independent envelope behavior region can be seen for each curve by inspection. The envelope fit curve, with extrapolation, is red} \protect\label{fig:1}
\end{figure}

\begin{figure}
  \includegraphics[width=3.5in]{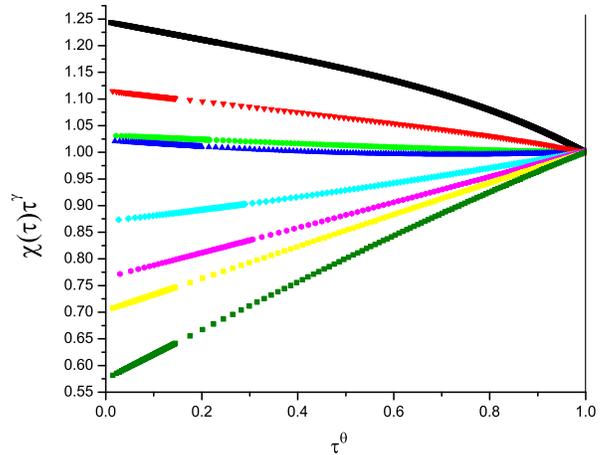}
  \caption{(Color online) The normalized reduced susceptibility
    $\chi(\tau)\tau^{\gamma}$ as a function of $\tau^{\theta}$ for the
    diamond, sc, bcc, fcc, $z18, z26, z32$ and $z56$ sc Ising models,
    top to bottom (black, red, green, blue, cyan, pink, yellow,
    olive). The exponents are taken to be $\gamma = 1.2371$ and
    $\theta =0.50$.  } \protect\label{fig:2}
\end{figure}
\begin{table}[htbp]
  \caption{\label{Table:I} Values of the fitting parameters for
    $\chi(\tau)$ in each of the models, as defined in
    Eq. \ref{chitau}.}
  \begin{ruledtabular}
    \begin{tabular}{cccccc}
      $z$&$\beta_c$&$C_{\chi}$&$a_{\chi}$&$k_{\chi}$&$\lambda_{\chi}$ \\
      $4$&$0.3697$&$1.245$&$-0.1382$&$-0.059$&$2$\\
      $6$&$0.221655$&$1.116$&$-0.0914$&$-0.0125$&$3$\\
      $8$&$0.15737$&$1.038$&$-0.073$&$0.036$&$1.6$\\
      $12$&$0.102067$&$1.022$&$-0.062$&$0.04$&$1.5$ \\
      $18$&$0.06442$&$0.871$&$0.111$&$0.038$&$0.9$\\
      $26$&$0.0430385$&$0.765$&$0.302$&$0.009$&$1$\\
      $32$&$0.0343267$&$0.703$&$0.428$&$-0.006$&$3$\\
      $56$&$0.0189291$&$0.575$&$0.800$&$-0.06$&$1.5$\\
    \end{tabular}
  \end{ruledtabular}
\end{table}
\begin{figure}
  \includegraphics[width=3.5in]{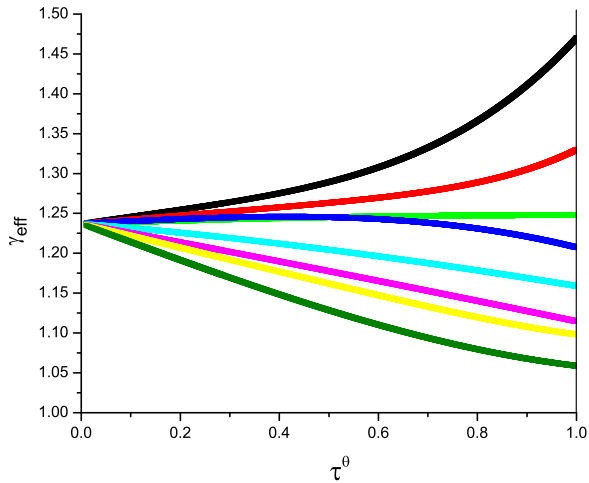}
  \caption{(Color online) The effective exponent
    $\gamma_{\mathrm{eff}}(\tau)$ as a function of $\tau^{\theta}$ for
    the diamond, sc, bcc, fcc, $z18, z26, z32$ and $z56$ sc Ising
    models, top to bottom (black, red, green, blue, cyan, pink,
    yellow, olive). } \protect\label{fig:3}
\end{figure}
The critical parameter estimates $\beta_c(z)$, $C_{\chi}(z)$ and
$a_{\chi}(z)$ and the approximate effective parameters $k_{\chi}(z)$
and $\lambda_{\chi}(z)$ are given in Table I. (The second
correction term is always weak, so the values of $\lambda_{\chi}(z)$ are not precise as they depend very sensitively on the fit parameters chosen for the other variables). All the models, including those with longer range interactions, follow the critical scaling rules up to infinite
temperature, with a gradual evolution of the critical amplitudes
as $z$ increases but without a trace of a crossover to mean-field
behavior at high $T$. It is important to note that it is the coordination number $z$ rather than the interaction range measured in terms of the nearest neighbor distance which is the key parameter (with weak lattice structure effects); the diamond, sc, bcc and fcc are all nearest neighbor lattices but they have significantly  different values for $C_{\chi}$ and $a_{\chi}$.

There seems no obvious reason to expect a
breakdown in these rules however large the range of interactions as
long as there is a cut-off so that the range remains finite; the correction amplitudes should continue to increase with increasing range. (If interactions fall off algebraically and sufficiently slowly, the
models will leave the finite-range universality class
\cite{fisher:72,suzuki:72}).

\begin{figure}
  \includegraphics[width=3.5in]{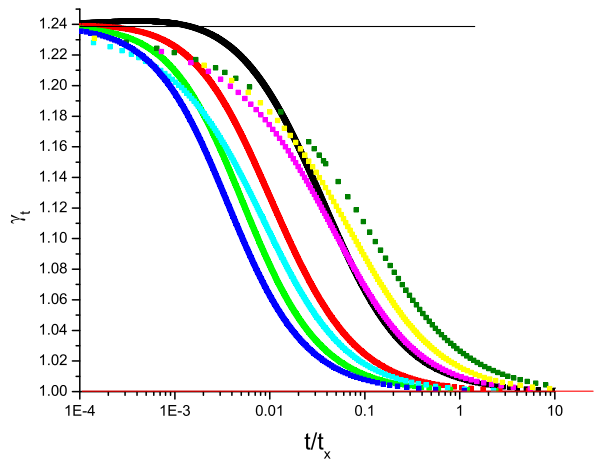}
  \caption{(Color online) The effective exponent
    $\gamma_{t}$ as a function of $t/t_{x}$ for
    the diamond, sc, bcc, fcc, $z18, z26, z32$ and $z56$ sc Ising
    models, (black, red, green, blue, cyan, pink,
    yellow, olive). $t_{x}$ is defined following Ref.~\cite{kim:03}.} \protect\label{fig:4}
\end{figure}

For comparison the data of Figure 3 translated appropriately are shown (Fig. 4) in the form of a $\gamma_{t}$ against $t/t_{x}$ where the normalization parameter is defined by $t_{x} = (a_{\chi,z})^{-1/\theta}$ as used in Ref.~\cite{kim:03}. It can be observed that there is no universality either in the position or the form of the individual curves.
\begin{figure}
  \includegraphics[width=3.5in]{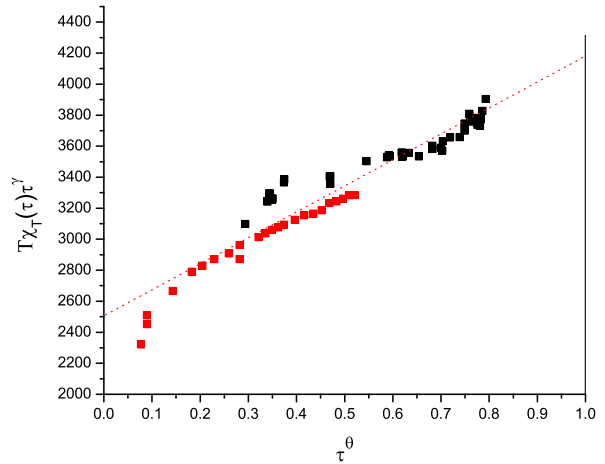}
  \caption{(Color online) The normalized reduced susceptibility
    $T\chi_{T}(\tau)\tau^{\gamma}$ of Ni as a function of $\tau^{\theta}$
    with the Heisenberg exponents $\gamma = 1.349$ and
    $\theta=0.55$. The experimental data and units are taken from
    \cite{weiss:26} (red circles) and \cite{fallot:44} (black
    squares). Following \cite{fallot:44} and \cite{souletie:83}, a
    small temperature independent term has been subtracted from the
    raw $\chi_{T}$ data.}  \protect\label{fig:5}
\end{figure}

Experiments can be interpreted using the same approach. The venerable
experimental data for the susceptibility of the ferromagnet Ni tabulated by Weiss and Forrer \cite{weiss:26} and by Fallot \cite{fallot:44} are exhibited in Fig. 5 in the same form as that used for the numerical data in Fig. 2. Here we consider Ni as a $S=1/2$ Heisenberg local moment system and so use the Heisenberg exponent values $\gamma= 1.396$ and $\theta =
0.55$ \cite{compostrini:02}. Following Fallot himself \cite{fallot:44}
and \cite{souletie:83}, we have subtracted out a small temperature
independent susceptibility term, which could well come from an orbital
contribution (see \cite{dupree:79} for the case of Co). As in the
Ising models with higher coordination numbers shown in Fig. 2, the
normalized reduced susceptibility increases almost linearly with
$\tau^\theta$ over the wide range of temperatures covered which
extends to $3T_c$, i.e. $\tau^{\theta} \sim 0.8$.  The ratio between
the asymptotic critical value of $\chi(\tau)\tau^{\gamma}$ and the
estimated extrapolated infinite temperature value (equal to $1$ for
spin $1/2$ in the appropriate units) can be taken as a measure of the
effective $C_{\chi}$. For Ni the observed ratio is about $0.60$, or alternatively the correction amplitude $a_{\chi} \sim 0.65$.
\begin{figure}
  \includegraphics[width=3.5in]{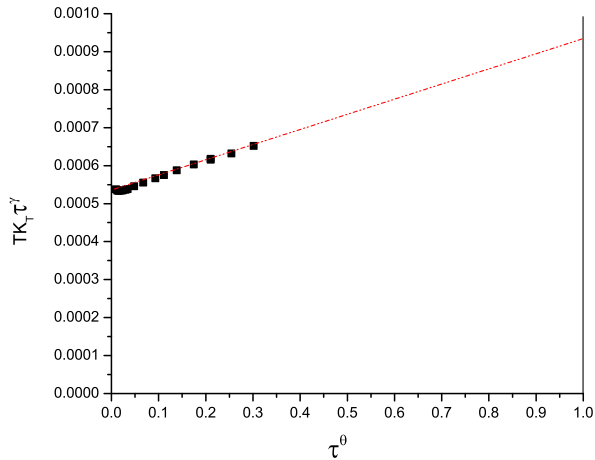}
  \caption{(Color online) The normalized reduced compressibility
    $TK_{T}\tau^{\gamma}$ of Xe as a function of
    $\tau^{\theta}$. Experimental data and units taken from
    \cite{guttinger:81}.}  \protect\label{fig:6}
\end{figure}

As an example of a gas-liquid transition we consider the susceptibility $\chi_{T}$ (defined as the derivative of the density by the chemical potential $\chi_{T} \equiv (\partial{\rho}/\partial{\mu})_{T}$) of Xe on the critical isochore for the liquid-gas transition, for which results from careful experiments based on light scattering techniques are tabulated
in Ref.~\cite{guttinger:81}. In Fig. 6 these data are plotted in the form
$T\chi_{T}\tau^{\gamma}$ against $\tau^{\theta}$, with the 3d Ising
exponents. (It can be noted that the susceptibility $\chi_{NN}$ defined by Orkoulas {\it et al} \cite{orkoulas:01} in their analysis of the hard-core square-well fluid is  $T\chi_{T}$; Orkoulas {\it et al} also use $\tau$ as the scaling variable). Again the figure shows an essentially linear increase of $T\chi_{T}\tau^{\gamma}$ with $\tau^{\theta}$ just as in the numerical plots for the Ising models with large coordination numbers. The ratio of the critical limit to the extrapolated high temperature limit is
$C_{\chi} \sim 0.55$, or alternatively the correction amplitude $a_{\chi} \sim 0.75$. This value is broadly consistent with the values $a_{\chi}= 1.3(2)$ \cite{guttinger:81} and $a_{\chi}=1.08$ \cite{anisimov:95} estimated from previous analyses based on the same data set but using different scaling rules. The complications associated with the asymmetry in the fluid phase diagram should be kept in mind, but this plot suggests that as in the magnetic case even if fluid data were available to much higher $\tau$ within the present approach there would be no need to invoke a crossover to mean-field like behavior.
\begin{figure}
  \includegraphics[width=3.5in]{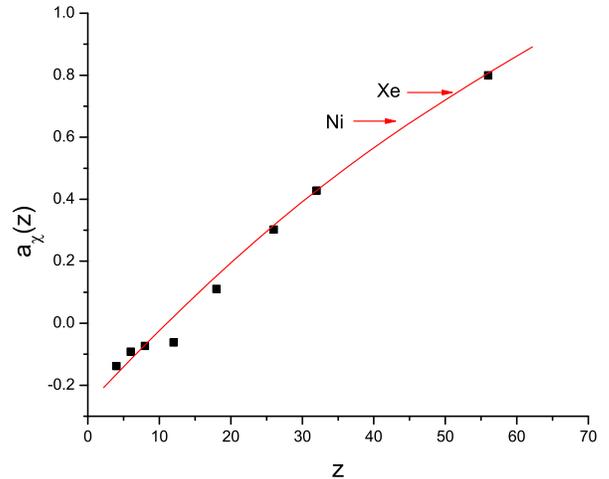}
  \caption{(Color online) The effective critical correction amplitudes
    $a_{\chi}(z)$ for the Ising models as a function of coordination number $z$, Table I, plus points for Ni and Xe obtained from the plots of the experimental data in Figs 5 and 6 (arrows).}
  \protect\label{fig:7}
\end{figure}
In Fig. 7 the values of $a_{\chi}$ for the numerical models are
plotted against $z$, and the effective values for Ni and for Xe are
indicated by arrows.  From this figure we can estimate the effective
coordination number $z$ for Ni and for Xe. To obtain a more quantitative estimate the Ni experimental data
should be compared to numerical results for Heisenberg
spins on an fcc lattice for different $z$ rather than for Ising spins
on an sc lattice. From data comparing $\gamma(\tau)$ on Heisenberg spins on   sc, bcc and fcc lattices \cite{fahnle:84} it appears that the numerical $a_{\chi}(z)$ plot
will be of similar form for Heisenberg spins as for Ising spins. One can then estimate that approximately $z \sim 45$, or in other words the effective interactions extend to between two and three lattice spacings. Obviously for real physical systems the model of equivalent interactions with cut-off is only a rough approximation to the true situation but the effective $z$ is a useful indicative phenomenological parameter.

For Xe the effective $a_{\chi}$ corresponds to an effective sc Ising
coordination number $z \sim 55$. When data for some other fluids are expressed graphically in terms of $\gamma_{eff}(\tau)$ \cite{anisimov:95}, $\gamma_{eff}$ initially increases slightly with increasing temperature,
meaning that $a_{\chi}$ is negative. (The experimental fluid data sets generally extend only over a narrow temperature range above $T_c$; for small $\tau$, $\gamma_{t}(t)$ is practically indistinguishable from $\gamma_{\tau}(\tau)$ and values estimated for $a_{\chi}$ are not sensitive to which scaling expression is used). Negative $a_{\chi}$ values have been observed in all aqueous electrolyte
solutions and also in many non-aqueous ionic solutions \cite{gutkowski:01}. It has been suggested \cite{anisimov:95} that as a general rule simple fluids have positive $a_{\chi}$ and complex fluids
negative $a_{\chi}$. A comparison with Table I and Fig. 3 indicates that for the negative $a_{\chi}$ systems the effective coordination number $z$ is $12$ or less while the positive $a_{\chi}$ systems have much higher effective $z$ values.

The effective coordination number is a fundamental parameter for
understanding the magnetism of metallic ferromagnets which can often
be considered either from a band or from a local moment
perspective. In the case of liquids, it should be possible to make a systematic classification of effective coordination numbers and to link these $z$ values to the interatomic potentials used for calculating structure functions.

In conclusion, the analysis given above leads to a simple overall physical scenario in which for a family of 3d Ising models the temperature dependence of the reduced susceptibility over the entire temperature range from $T_c$ right up to infinite temperature is explained using the critical RGT formalism with appropriate Wegner corrections and without the need to invoke a restricted "critical region" or any form of high temperature crossover. The approach is conceptually economical and leads to a transparent interpretation of the differences in behavior from model to model and from system to system within a universality class; there is a strong correlation between the
coordination number and the strength of the non-universal amplitude of the leading confluent Wegner term, which dominates the corrections.

We acknowledge gratefully an interesting discussion of the
experimental high temperature effective exponents with Ralph
Chamberlin, and helpful explanations on the thermodynamics of fluids from Jan Sengers. This research was conducted using the resources of High
Performance Computing Center North (HPC2N).

\end{document}